\documentclass[conference]{IEEEtran}
\IEEEoverridecommandlockouts
\usepackage{cite}
\usepackage{amsmath,amssymb,amsfonts}
\usepackage{algorithmic}
\usepackage{graphicx}
\usepackage{textcomp}
\usepackage{xcolor}
\usepackage{bm}
\usepackage{hyperref}
\usepackage{threeparttable}
\usepackage{cite}
\usepackage{makecell}

\def\BibTeX{{\rm B\kern-.05em{\sc i\kern-.025em b}\kern-.08em
    T\kern-.1667em\lower.7ex\hbox{E}\kern-.125emX}}
\begin{document}

\title{A Sparse Polynomial Chaos Expansion-Based Method for Probabilistic Transient Stability Assessment and Enhancement

\thanks{This work was supported by Natural Sciences  and Engineering Research Council (NSERC) Discovery Grant, NSERC RGPIN-2016-04570.}
}

\author{Jingyu Liu,~\IEEEmembership{Student Member,~IEEE}, Xiaoting Wang,~\IEEEmembership{Student Member,~IEEE}, Xiaozhe Wang,~\IEEEmembership{Senior Member,~IEEE}}

\maketitle

\begin{abstract}
This paper proposes an adaptive sparse polynomial chaos expansion(PCE)-based method to quantify the impacts of uncertainties on critical clearing time (CCT) that is an important index in transient stability analysis. 
The proposed method can not only give fast and accurate estimations for the probabilistic characteristics (e.g., mean, variance, probability density function) of the probabilistic CCT (PCCT), but also provides crucial information about the sensitivity of random inputs with respect to the variance of PCCT. Utilizing the sensitivity information, mitigation measures can be developed for transient stability enhancement.  
Numerical studies on the WSCC 9-bus system demonstrate the high accuracy and efficiency of the proposed method compared to the Monte Carlo simulation method. The provided sensitivity information and the effectiveness of mitigation measures in transient stability enhancement are also verified. 
\end{abstract}


\section{Introduction}

The increasing incorporation of renewable energy sources (RES) and new forms of load demand is introducing growing uncertainties into power girds, the influence of which on transient stability needs to be studied carefully. Particularly, the critical clearing time (CCT) that is typically used as a crucial index in transient stability assessment 
is no longer a deterministic value. Rather, it becomes a random variable when various uncertainty sources are considered. The transient stability needs to be assessed in a probability sense. 

To address the challenges of uncertainties, 
Monte Carlo simulations (MCS) based on time-domain simulation (TDS) have been adopted in \cite{faried2010} to assess the statistics of the probabilistic critical clearing time (PCCT). However, MCS is computationally expensive and may not be feasible for large-scale systems in practical applications. 
To improve the efficiency, sensitivity-based methods, transient energy function (TEF)-based methods and their combination are employed in 
\cite{2019Fortulan,Roberts2015, taj2018}. However, assumptions and simplifications (e.g., constant impedance assumption) are inevitable when deriving TEF, which may affect the computation accuracy. 
In addition, emerging machine learning (ML) techniques have been adopted for CCT related online transient stability prediction \color{black} \cite{Songkai2020,zhu2019}. 
Despite their accurate pointwise estimations of CCT, 
the impacts of various uncertainty sources on CCT were not investigated while sensitivity information crucial for transient stability enhancement cannot be obtained. 

Recently, a metamodeling method called polynomial chaos expansion (PCE) has been applied in solving probabilistic transfer capability\cite{Sheng2018, Wang2021}, stochastic economic dispatch \cite{Wang2021ED}, \color{black}and probabilistic transient stability problems\cite{ Bingqing2019, Yijun2019,Miao2021}. Using only a small number of input samples, the PCE-based methods can generate  surrogate models whose embedded coefficients explicitly interpret accurate statistics and sensitivity information of the system response.
However, generator rotor angle trajectories are the responses of interest in \cite{ Bingqing2019, Yijun2019,Miao2021}, which may not quantify the stability margin or probability of stability directly. Besides, control measures to mitigate the negative impacts of uncertainty on power system transient stability remain to be seen. 

In this paper, we propose an adaptive sparse PCE-based method for probabilistic transient stability assessment and enhancement. The proposed method can assess the probabilistic characteristics of PCCT (e.g., mean, variance, probability density function (PDF)) accurately and efficiently while pinpointing the most sensitive random inputs with respect to the variation of PCCT. Leveraging the provided sensitivity information, mitigation measures to reduce the variation of PCCT is developed for the enhancement of transient stability.  

\section{Probabilistic Critical Clearing Time}
The traditional electromechanical transient model of power system can be formulated as a set of differential  algebraic equations (DAEs):

\begin{small}
\vspace{-0.1cm}
\begin{equation}
\begin{aligned}
\dot{\bm{x}} &= \bm{f}(\bm{x}, \bm{y}, \bm{\lambda}, \bm{u})  \\
\bm{0} &= \bm{h}(\bm{x}, \bm{y}, \bm{\lambda}, \bm{u})
\label{eq:PSDAE}    
\end{aligned}
\end{equation}
\vspace{-0.3cm}
\end{small}

\noindent where $\bm{f}$ are differential equations (e.g., swing equations); $\bm{h}$ are algebraic equations (e.g., power flow); $\bm{x}$ are state variables (e.g., rotor speeds and rotor angles of generators); $\bm{y}$ are algebraic variables (e.g., bus voltage magnitudes and angles); $\bm{\lambda}$ are system parameters (e.g., 
load power); $\bm{u}$ are discrete variables modeling events (e.g., fault occurrence and switching operation of tap-changers).

\begin{figure}[!t]
\centering
\includegraphics[width=0.8\columnwidth]{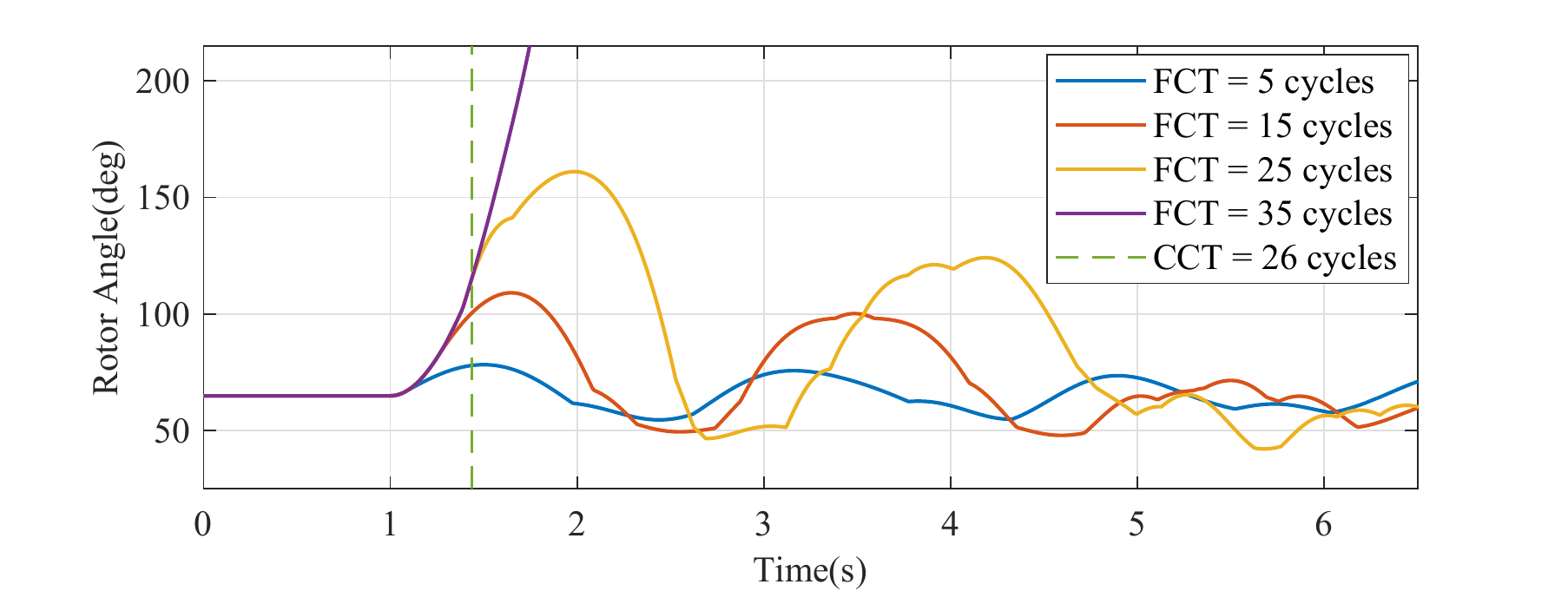}
\vspace{-0.3cm}
\caption{Illustration of maximum rotor angle difference trajectories. A fault is applied at $t = 1.0$s, and is cleared after different fault clearing time (FCT).}
\label{fig:CCTExplanation}
\vspace{-0.5cm}
\end{figure}

Fig. \ref{fig:CCTExplanation} presents an example of the maximum rotor angle difference trajectories after a fault occurred at $1.0$s, which are  obtained by numerical integrating \eqref{eq:PSDAE}. 
In Fig. \ref{fig:CCTExplanation}, fault clearing time (FCT) $T$ represents the time duration between fault occurrence and clearance. Increasing $T$ typically causes system instability (e.g, the purple solid line indicating a diverged trajectory in Fig. \ref{fig:CCTExplanation}). The maximum value of $T$ that maintains system stability is the CCT ($T_{\mathrm{cct}}$). $T_{\mathrm{cct}} - T$ is a well-accepted stability margin measure in transient stability studies\cite{zhu2019}.


In deterministic cases, $T_{\mathrm{cct}}$ is a constant determined by preset $\bm{\lambda}$ and $\bm{u}$ (e.g., fault locations). When RES are integrated in the system, parts of $\bm{\lambda}$ are modelled as random variables $\bm{\chi}$ (e.g., Gaussian distribution for load variations) and the others $\bm{\lambda}^{\prime}$ remain preset constants. Therefore, system \eqref{eq:PSDAE} can be reshaped as follows:

\begin{small}
\vspace{-0.1cm}
\begin{equation}
\begin{aligned}
         \dot{\bm{x}} &= \bm{f}(\bm{x}, \bm{y}, \bm{\lambda}^{\prime}, \bm{\chi},  \bm{u})  \\
           \bm{0} &= \bm{h}(\bm{x}, \bm{y}, \bm{\lambda}^{\prime}, \bm{\chi}, \bm{u})
\label{eq:Stochastic-PSDAE}        
\end{aligned}
\end{equation}
\vspace{-0.3cm}
\end{small}

\noindent and the resulting CCT becomes a random variable $T_{\mathrm{pcct}}$, i.e., PCCT.
The traditional way to estimate the probabilistic characteristics of the PCCT is to run the MCS using TDS. This method utilizes binary search strategy \cite{zhu2019} to compute exact CCT for every realization of $\bm{\chi}$. Then, important statistics of PCCT (e.g., mean, variance, PDF, and cumulative density  function (CDF)) and probability of stability can be obtained. However, MCS requires expensive computation effort. To 
enhance efficiency, a PCE-based method is implemented in this study to estimate the probabilistic characteristics of the PCCT through only a small number of scenarios. Important byproduct information of sensitivity analysis can also be generated after the PCE-based model is constructed, which is further used to design control measures for transient stability enhancement. 

\section{Adaptive Sparse Polynomial Chaos Expansion}\label{section_3}
\subsection{Polynomial Chaos Expansion}\label{sec:PCE_construction}
%
Let $\bm{\chi}=\left[\chi_1, \chi_2,...,\chi_M\right]$ be an $M$-dimensional independent random inputs  (e.g., load variations). 
According to the polynomial chaos theory introduced by \cite{wiener1938homogeneous}, $T_{\mathrm{pcct}}$ that can be solved from the transient stability model in \eqref{eq:Stochastic-PSDAE} can be expressed by a weighted sum of orthogonal polynomials:

\begin{small}
\vspace{-0.1cm}
\begin{equation}
T_{\mathrm{pcct}} =g(\bm{\chi}) \approx \sum_{i=1}^{L} a_i \Psi_i(\bm{\chi})
\label{eq:PCE_formula}
\end{equation}
\end{small}

\noindent where ${\big\Vert{T_{\mathrm{pcct}} - \sum_{i=1}^{L} a_i \Psi_i(\bm{\chi}) }\big\Vert}_{\mathcal{L}_2}^{2} \to 0$ as the number of expansion terms $L \to \infty$. $a_i$ is the $i$-th expansion coefficient to be determined and $\Psi_i(\bm{\chi})$ denotes the corresponding multidimensional orthogonal polynomial basis with respect to the joint PDF $f_{\bm{\chi}}(\bm{\chi})$. \color{black}Generally, the polynomial basis $\Psi_{i}(\bm{\chi})$ can be constructed as the tensor product of their univariate counterparts: 

\begin{small}
\vspace{-0.4cm}
\begin{equation}
     \Psi_{i}(\bm{\chi}) = \prod_{j=1}^M \psi^{(j)}_{k_{ij}}(\chi_j)
 \label{eq:multivariatebasis}
\end{equation}
\vspace{-0.3cm}
\end{small}

\noindent where $\psi^{(j)}_{k_{ij}}(\chi_j)$ refers to the $k_{ij}$-th order univariate polynomial basis of $\chi_j$ at $i$-th expansion term.
For $\chi_j$ with given typical continuous distributions, the polynomials $\psi^{(j)}$ can be selected from Table \ref{tab:WA-Poly}. For other non-typical distributions with finite moments, $\psi^{(j)}$ can be numerically built based on the Stieltjes procedure \cite{UQdoc_PCE}. 
 
%
\begin{table}[htbp]
\vspace{-0.5cm}
\setlength{\abovecaptionskip}{-0cm}
\setlength{\belowcaptionskip}{-0cm}
\caption{Correspondence between standard continuous distributions and families of orthogonal polynomials \cite{Blatman2009}}
\label{tab:WA-Poly}
\centering
\begin{tabular}{c|c|c}
\hline
Distribution & Polynomial & Support \\ \hline
Gaussian     & Hermite    & $\mathbb{R}$     \\\hline
Uniform      & Legendre   &   $[-1, 1]$      \\\hline
Gamma        & Laguerre   &   $(0, +\infty)$      \\\hline
Beta         & Jacobi     &   $(-1, 1)$      \\ \hline
\end{tabular}
\vspace{-0.5cm}
\end{table}

\subsection{The Adaptive Sparse Procedure}\label{subsection_3b}
\noindent $\bm{q}$\textbf{-Norm Truncation Scheme}: $\Psi_{i}(\bm{\chi})$ is typically truncated by setting a total-degree limit $p$,  
i.e., $\sum_{j=1}^M k_{ij}\leq p$,  
leading to a total number of expansion terms $L=(M+p)!/(M!p!)$ that increases  exponentially  with  both $p$ and $M$. To reduce $L$, 
we adopt the $q$-norm adaptive sparse truncation scheme \cite{Blatman2009} based on the fact that most models are governed by the main effects and lower-order interactions \cite{Blatman2009}. 
As such, the degree indices $\bm{{k}}_{i} = \lbrace k_{i1},...,k_{iM} \rbrace$ of the retained $\Psi_i(\bm{\chi})$ 
must 
\color{black} satisfy:

\begin{small}
\vspace{-0.2cm}
\begin{equation}
\|\bm{{k}}_{i}\|_{q}\leq p,\  q\in(0,1)
\label{eq:qNormTrunc}
\end{equation}
\vspace{-0.5cm}
\end{small}

\noindent where $p$ and $q$ in \eqref{eq:qNormTrunc} can be chosen adaptively based on the modified leave-one-out (MLOO) cross-validation error ((1.22) in \cite{UQdoc_PCE}). 

\noindent \textbf{
Hybrid-Least Angle Regression (LAR)\color{black}}:
A combination of LAR and Ordinary Least Square (OLS), is employed to calculate the coefficients $a_i$. Specially, LAR is used to select the optimal basis polynomials  $\Psi_i(\bm{\chi})$ that are most relevant to the model response $T_{\mathrm{pcct}}$, from the retained basis set determined by \eqref{eq:qNormTrunc}, whereas OLS is used for coefficients calculation and quick MLOO error evaluation by solving the following minimization problem:

\begin{small}
\vspace{-0.2cm}
\begin{equation}
 \setlength{\abovedisplayskip}{2pt}
 \setlength{\belowdisplayskip}{2pt}
\label{eq:minimize_pro}
\begin{aligned}
        \min \sum_{l =1}^{N}\left[ T_\mathrm{pcct}^{(l)} - \sum_{i=1}^{L}a_i \Psi_i(\bm{\chi}^{(l)})\right]^2 
\end{aligned}
\end{equation}
\vspace{-0.2cm}
\end{small}

\noindent where $(\bm{\chi}^{(l)}, T_\mathrm{pcct}^{(l)})$, $l=1,...,N$ are sample pairs computed by the TDS method.
Detailed formulation of Hybrid-LAR can be found in \cite{UQdoc_PCE}. 
 The Hybrid-LAR algorithm overcomes the limitations of LAR and OLS, which can solve underdetermined problems 
efficiently and enhance the sparsity of PCE-based models, while maintaining good estimation accuracy.


\subsection{Estimating the Probabilistic Characteristics of PCCT}
\noindent \textbf{Response Estimation}: Since the basis polynomial $ \Psi_i(\bm{\chi})$
and expansion coefficients $a_i$ have been determined, the \color{black} well constructed PCE-based model (\ref{eq:PCE_formula}) can be directly utilized as a surrogate model. The model responses and their PDF 
can be estimated with good accuracy and cheap computation cost by substituting a large number of scenarios of random inputs to the surrogate model. 

\noindent \textbf{Moments of PCE}: The orthonormality of basis polynomials induces PCE coefficients encoded important information of the response moments, i.e., mean and variance. If $ \Psi_1(\bm{\chi}) = 1$:

\begin{small}
\vspace{-0.2cm}
\begin{equation}
 \setlength{\abovedisplayskip}{2pt}
 \setlength{\belowdisplayskip}{2pt}
\mu(T_{\mathrm{pcct}})\approx\mathbb{E}\left[ \sum_{i=1}^{L} a_i \Psi_i(\bm{\chi}) \right] = a_1
\label{PCEMean} 
\end{equation}
\begin{equation}
 \setlength{\abovedisplayskip}{2pt}
 \setlength{\belowdisplayskip}{2pt}
\sigma^2(T_{\mathrm{pcct}})\approx\mathrm{Var}\left[ \sum_{i=1}^{L} a_i \Psi_i(\bm{\chi}) \right] =  \sum_{i=2}^L a_i^2
\label{PCEVar}
\end{equation}
\vspace{-0.3cm}
\end{small}

\subsection{PCE-Based Sensitivity Analysis}
\normalsize
The orthonormality of basis polynomials also builds a connection between the coefficients $a_i$ and Sobol' indices (a.k.a Analysis Of Variance, ANOVA) that can be used to pinpoint the most critical random inputs to the variation of the system response. Such information is critical to alleviate the negative impacts of uncertainty on the system performance. 
Sobol' indices are defined based on so-called Sobol' decomposition \cite{Sobol1993}. For integrable function $g(\bm{\chi})$ with independent random inputs $\bm{\chi}$, there exists a unique expansion:

\begin{small}
\vspace{-0.1cm}
\begin{equation}
g(\bm{\chi}) = g_0 + \sum_{\substack{\bm{s} \subseteq \lbrace 1,...,M \rbrace \\ \bm{s}\not = \emptyset}}g_{\bm{s}}(\chi_{\bm{s}}) 
\label{eq:SobolDecom}
\end{equation}
\vspace{-0.3cm}
\end{small}

\noindent where $g_0$ denotes the mean value of $g(\bm{\chi})$; $\chi_{\bm{s}}$ represents a subset of $\bm{\chi}$ depending on the index $\bm{s}$ (e.g., for $\bm{s}=\lbrace 1,2 \rbrace$, $\chi_{\bm{s}}=\lbrace \chi_1,\chi_2 \rbrace$); $g_{\bm{s}}(\chi_{\bm{s}})$ is a function that only depends on $\chi_{\bm{s}}$, which satisfies  
$\int_{\Omega_j}g_{\bm{s}}(\chi_{\bm{s}})f_{\chi_j}(\chi_j)d{\chi_j} = 0,\ \forall j\in\bm{s}$. 
%
Introducing an index set $
\bm{I}_{\bm{s}} = \{ i\in \{ 1,...,L \}: k_{ij}\not= 0\Leftrightarrow j\in \bm{s} \}$, \color{black} 
the Sobol' terms in \eqref{eq:SobolDecom} can be approximated by the PCE-based model:

\begin{small}
\vspace{-0.2cm}
\begin{equation}
g_{\bm{s}}(\chi_{\bm{s}}) \approx \sum_{i\in \bm{I}_{\bm{s}}} a_i \Psi_i(\bm{\chi})
\label{eq:PCE_Sobol_approx}
\end{equation}
\vspace{-0.25cm}
\end{small}

\noindent By \eqref{PCEVar}, the two important Sobol' indices, the first-order indices $S_{\zeta}^{(1)}$ \color{black} and the total indices $S_{\zeta}^{(T)}$ with $\zeta=1,...,M$ can be approximated by the coefficients of PCE-based model  (\ref{eq:PCE_formula}):\color{black}
%

\begin{small}
\vspace{-0.1cm}
\begin{equation}
S_{\zeta}^{(1)}=\frac{\mathrm{Var}[g_{\zeta}(\chi_{\zeta})]}{\mathrm{Var}[g(\bm{\chi})]}
\approx
\frac{\sum_{i\in \bm{I}_{\zeta}} a_i^2}{\sum_{i=2}^L a_i^2}
\label{eq:FirstOrderSI}
\end{equation}

\begin{equation}
S_{\zeta}^{(T)} = \frac{\mathrm{Var}\left[\sum\limits_{{\bm{s}\in\bm{s}_{\zeta}}}g_{\bm{s}}(\chi_{\bm{s}})\right] }{\mathrm{Var}[g(\bm{\chi})]} 
\approx 
\frac{\sum \limits_{\bm{s}\in\bm{s}_{\zeta}}\sum\limits_{{i\in \bm{I}_{\bm{s}}}}a_i^2}{\sum_{i=2}^L a_i^2}
\label{eq:TotalSobolSI}
\end{equation}
\vspace{-0.15cm}
\end{small}

\noindent where $ \bm{I}_{\zeta} = \{ i\in \{ 1,...,L\}: k_{ij}\not= 0\Leftrightarrow j=\zeta \}$ and $ \bm{s}_{\zeta} = \{\bm{s} \subseteq \{1,...,M\}:\zeta \in \bm{s}\}$.
The Sobol' indices can quantify the variance contribution of certain inputs $\chi_{\zeta}$ to the variance of system output $T_{\mathrm{pcct}}$. $S_{\zeta}^{(1)}$ describes the main contribution of $\chi_{\zeta}$ to the variance of $T_{\mathrm{pcct}}$, while $S_{\zeta}^{(T)}$ describes the contribution of $\chi_{\zeta}$ including its interactions with other inputs to the variance of $T_{\mathrm{pcct}}$\cite{Sobol1993}. 
Obviously, $S_{\zeta}^{(1)}$ or $S_{\zeta}^{(T)}$  with the highest values indicate the largest $T_{\mathrm{pcct}}$ variation share from the input $\chi_{\zeta}$, based on which the critical inputs $\chi_{\zeta}$ are found. 
With this information, the variation of the system response $T_{\mathrm{pcct}}$ can be effectively alleviated by smoothing out the dominant  inputs $\chi_{\zeta}$. 
The PCE-based Sobol' indices estimation can achieve good accuracy with far less computational cost  by using  (\ref{eq:FirstOrderSI})-(\ref{eq:TotalSobolSI}) directly, compared with the traditional Monte Carlo-based  Sobol' indices estimation \cite{UQdoc_Sen}. 

\section{The PCE-based algorithm for PCCT assessment and variation control} 
\begin{figure}[!t]
\centering
\includegraphics[width=0.75\columnwidth]{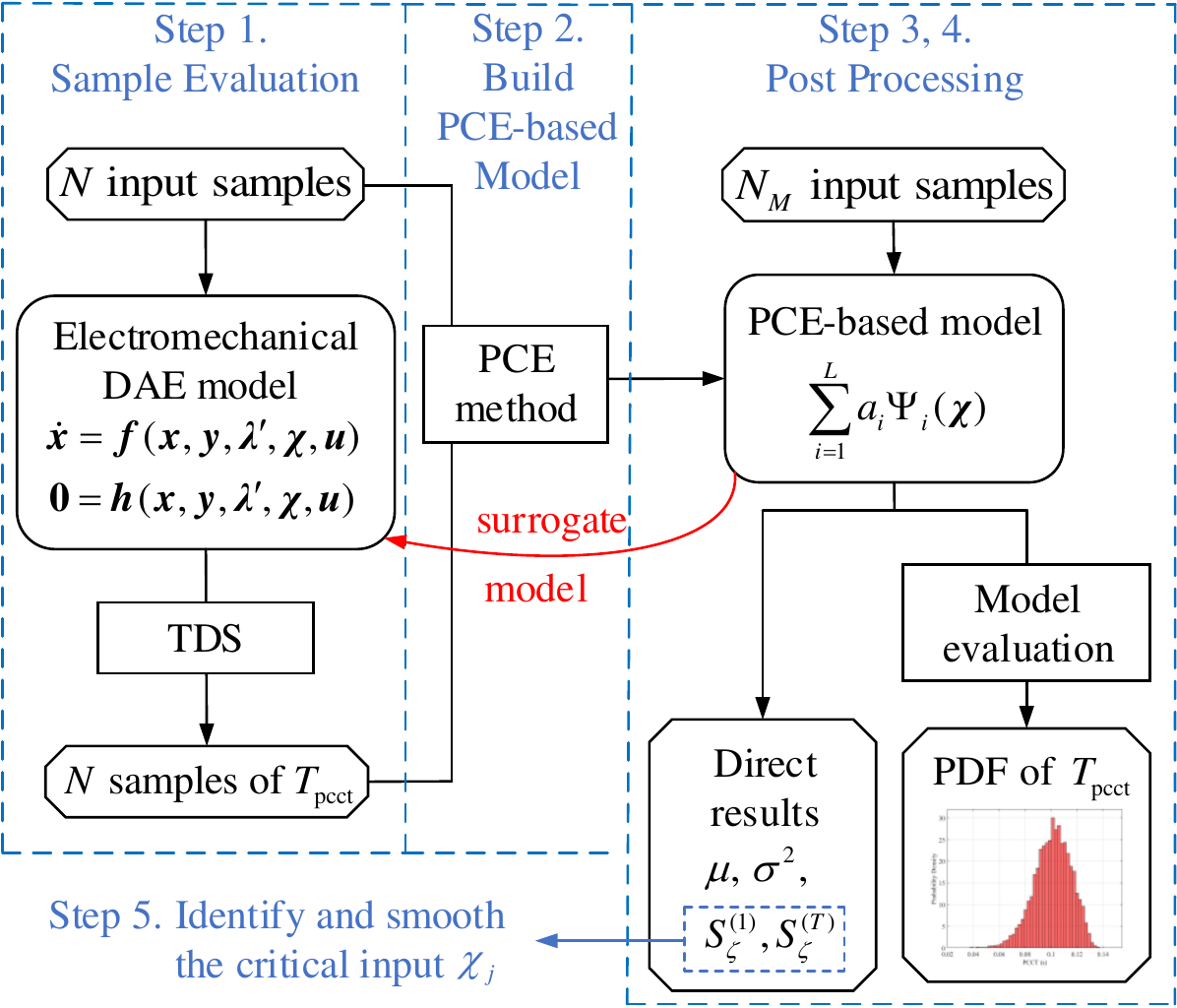}
\vspace{-0.2cm}
\caption{PCCT assessment and sensitivity analysis by PCE method.}
\label{fig:algorithmn}
\vspace{-0.6cm}
\end{figure}

Fig. \ref{fig:algorithmn} provides an overview of the computation procedure, which can be divided into five steps. 

\noindent\textbf{Step 1.} 
Generate $N$ realizations of random inputs $\bm{\chi}$ ( e.g., renewable generator outputs, load power) by Latin hypercube sampling (LHS) method. Evaluate the corresponding 
$T_{\mathrm{pcct}}$ through the TDS of \eqref{eq:Stochastic-PSDAE}.

\noindent\textbf{Step 2.} 
Build univariate basis polynomials $\psi^{(j)}$ according to Table \ref{tab:WA-Poly} or by Stieltjes procedure \cite{UQdoc_PCE}. Form the multivariate basis polynomials $\Psi_{i}(\bm{\chi})$ by (\ref{eq:multivariatebasis}) and apply the $q$-norm truncation scheme (\ref{eq:qNormTrunc}). Calculate coefficients by solving \eqref{eq:minimize_pro} using the Hybrid-LAR algorithm based on the $N$ sample pairs $(\bm{\chi}^{(l)}, T_\mathrm{pcct}^{(l)})$, $l=1,...,N$ obtained in \textbf{Step 1}.


\noindent\textbf{Step 3.} 
Compute mean, variance and Sobol' indices of PCCT using \eqref{PCEMean}-\eqref{PCEVar} and \eqref{eq:FirstOrderSI}-\eqref{eq:TotalSobolSI}. 

\noindent\textbf{Step 4.} Generate a large number of $N_M$ input samples ($N_M\gg N$) and evaluate their $T_{\mathrm{pcct}}$ using the established PCE-based model. From the estimated PDF and CDF of the $T_{\mathrm{pcct}}$, determine the PCCT with a desired confidential level $P_{cl}\%$ or calculate the probability of transient stability for a given FCT. 



\noindent\textbf{Step 5.}  Identify the critical random input dominating the variation of $T_\mathrm{pcct}$ from the the obtained first-order and total Sobol' indices. Given the available resources (e.g., the number and location of energy storage systems), smoothing out the top critical random inputs to reduce the variation of the PCCT and enhance the transient stability. 
\color{black}
\section{Case Studies}
\begin{figure}[!b]
\vspace{-0.4cm}
\centering
\includegraphics[width=0.75\columnwidth]{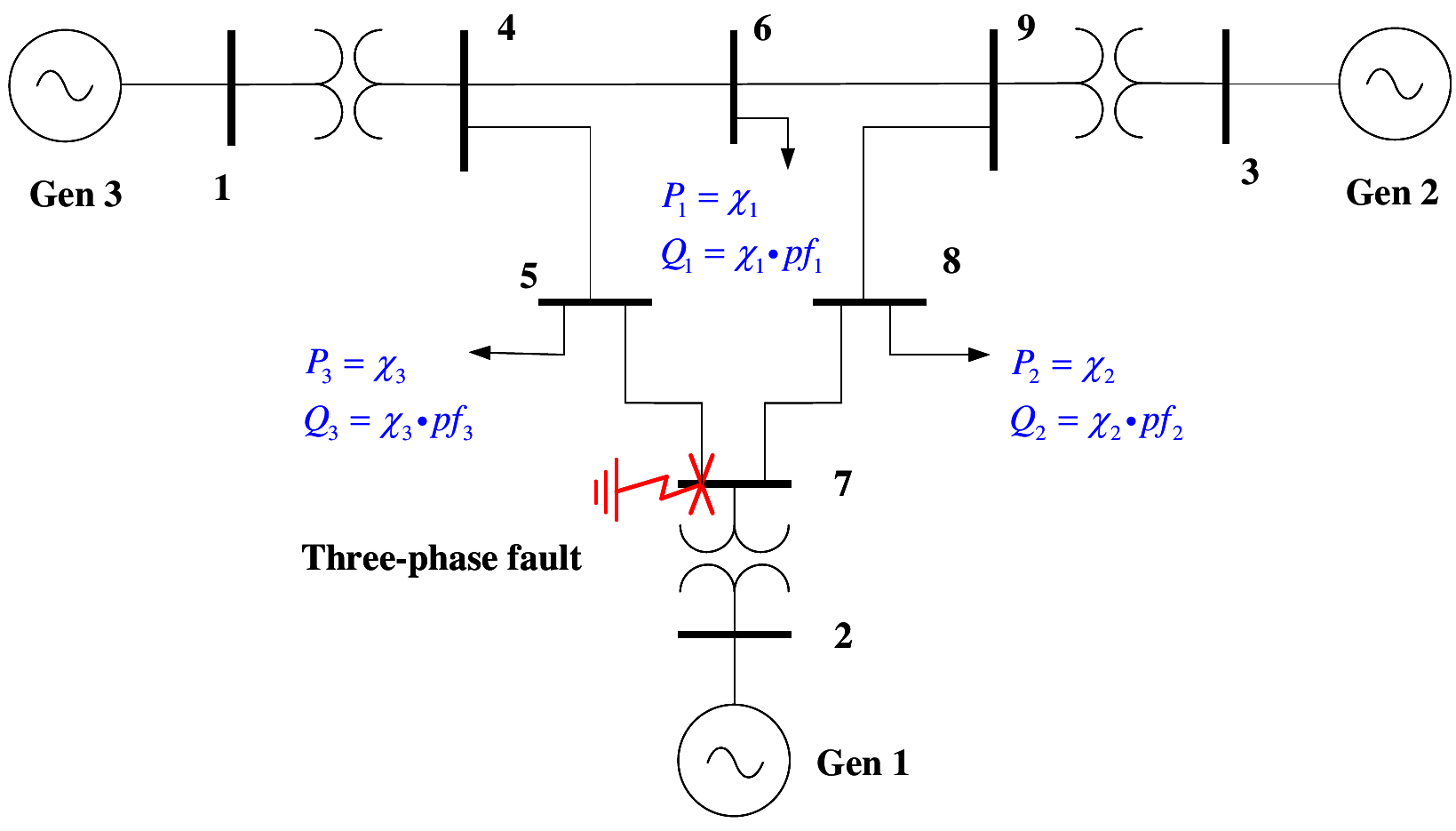}
\vspace{-0.2cm}
\caption{WSCC 3-generator 9-bus test system. Random loads are marked as blue and $pf_j$ are the power factors from original deterministic system.}
\label{fig:9BusSystem}
\end{figure}

In this section, the proposed method is tested in the WSCC 3-generator 9-bus system shown in Fig. \ref{fig:9BusSystem}.

The fourth order generator model, type \uppercase\expandafter{\romannumeral2} turbine governor (TG) model and the IEEET1 automatic voltage regulator (AVR) model in \cite{Milano2005} are implemented in PSAT \cite{Milano2005} for calculating the CCT through the TDS. 
It is assumed that all the three constant PQ loads have active power consumption following Gaussian distributions with mean values equal to the original deterministic values and standard deviations equal to $10\%$ of their means. This assumption is for illustration purpose. The proposed PCE  method is not restricted by the distribution types 
of random inputs. 
A constant power factor remains when load power varies.  A three-phase ground fault is applied at bus 7 at $1.0$s and is cleared after 5 cycles, i.e., $0.083$s, by opening line 5-7.  
The duration of TDS after fault clearance is set as $10$s. The computations are performed in MATLAB R2019b on an Intel Core i5-8500 CPU 3.00GHz CPU. 
The UQLab\cite{marelli2014} toolbox is used to construct the PCE-based model. 


 

\vspace{-0.1cm}
\subsection{Probabilistic Characteristics of PCCT} \label{sec:Stastics_PCCT}
First, $N$ = 30 samples of load power $\bm{\chi}$ are generated by LHS method 
and the corresponding $T_{\mathrm {pcct}}$ are evaluated via TDS of \eqref{eq:Stochastic-PSDAE} in PSAT. Based on the evaluated $N$ sample pairs of ($\bm{\chi},T_{\mathrm{pcct}}$), the PCE method in \textbf{Step 2} generates a PCE-based model \eqref{eq:PCE_formula} with 16 polynomial terms. Then the mean and variance of the PCCT can be quickly evaluated as in \textbf{Step 3}.
MCS results based on $10^4$ input samples are used as benchmark to demonstrate the performance of proposed PCE-based method. Table \ref{tab:WSCC9_stat_comp} lists the mean and variance of PCCT estimated by MCS and PCE, from which we can see that PCE can estimate the mean and variance of PCCT accurately.

%
\begin{table}[htbp]
\vspace{-0.5cm}
\setlength{\abovecaptionskip}{-0cm}
\setlength{\belowcaptionskip}{-0cm}
\renewcommand{\arraystretch}{1.0}
\caption{Comparison of estimated statistics and probability of stability}
\label{tab:WSCC9_stat_comp}
\centering
\scalebox{0.88}{
\begin{tabular}{c|c|c||c|c||c|c}
\hline
Method & ${\mu}$ &
${\frac{\Delta \mu}{\mu_{\mathrm{mcs}}}}$ &
${\sigma^2(10^{-4})}$ &
${\frac{\Delta \sigma^2}{\sigma_{\mathrm{mcs}}^2}}$  &
${P^{(S)}}$ &
${\frac{\Delta P^{(S)}}{P^{(S)}_{\mathrm{mcs}}}}$
\\ [6pt]
\hline
MCS    & $0.1013$  &  $0$    & $2.055$  & $0$ & $89.25\%$ & $0$ \\
\hline
PCE    & $0.1011$  &  $0.16\%$    & $2.004$ &  $2.49\%$ & $88.99\%$ & $0.29\%$ \\
\hline 
\end{tabular}
}
\begin{tablenotes}
\item * $P^{(S)}$ denotes the probability of stability under preset FCT $=5$ cycles.
\end{tablenotes}
\vspace{-0.3cm}
\end{table} 

\begin{figure}[!t]
\centering
\includegraphics[width=0.75\columnwidth]{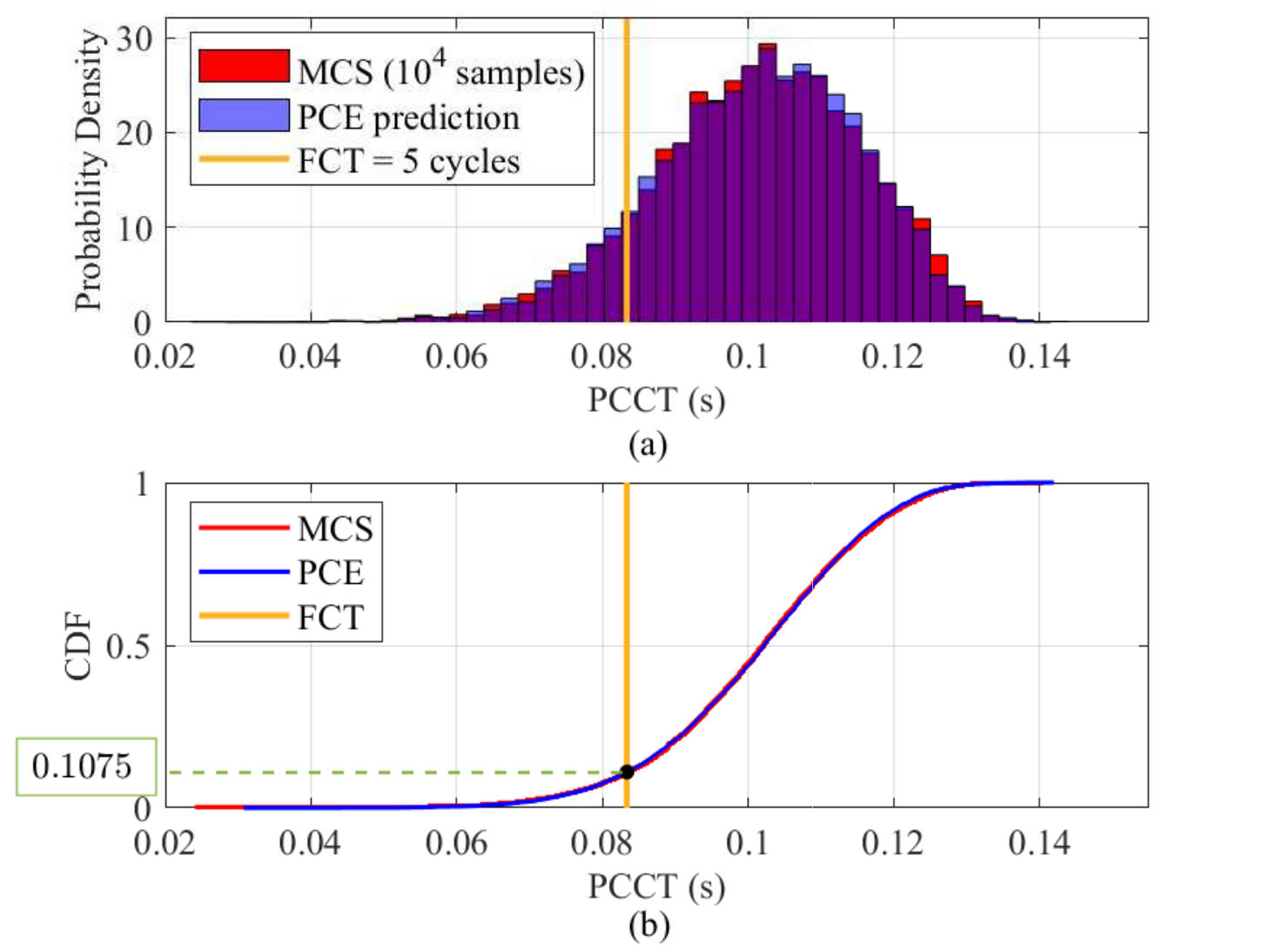}
\vspace{-0.2cm}
\caption{\color{black}A comparison of the PDF and CDF of PCCT obtained from MCS and those obtained by the proposed PCE-based model. The purple part shows that they are almost overlapped. If FCT is preset to be 5 cycles, the probability of the system to maintain transient stability is $89.25\%$.} 
\label{fig:PDFOfPCCT}
\vspace{-0.5cm}
\end{figure}

To assess the PDF and CDF of the PCCT,  $N_M = 10^4$  samples are generated and their PCCT values are calculated using the PCE-based model in \textbf{Step 4}. The results and the corresponding comparisons with MCS are shown in 
Fig. \ref{fig:PDFOfPCCT}, showing good estimations of the PDF and CDF of PCCT. The CDF of PCCT is further used to determine the
desired value of PCCT with a certain probability. For example, if the FCT is 5 cycles, i.e., $0.083$s, the probability of system to maintain stability after the fault is $89.25\%$ (i.e., $100\%-10.75\%$).  To ensure transient stability with $95\%$ confidence level, FCT needs to be $0.076$s (obtained from the 5th percentile $P_{5\%}$ of $T_{\mathrm{pcct}}$).  

Regarding the efficiency, Table \ref{tab:WSCC9_time_comp} compares the computational time needed for the MCS and the PCE, where $t_{\mathrm{ed}}$ is the time for calculating $N = 30$ sample pairs $(\bm{\chi}^{(l)}, T_\mathrm{pcct}^{(l)})$ using TDS; $t_{\mathrm{pc}}$ is the time to construct the PCE-based model; $t_{\mathrm{es}}$ is the time for evaluating $N_M = 10^4$ samples; and $t_{\mathrm{total}}$ is the total time required. It can be seen that once the small sample evaluation is done in \textbf{Step 1}, constructing the PCE-based model in \textbf{Step 2} and sample evaluation in \textbf{Step 4} takes negligible time. 
In this case, the proposed PCE-based method is $320$ times more efficient than MCS. The reason for large $t_{es}$ needed by the MCS is that each PCCT evaluation 
requires in average $14.8$ runs of complete TDS or $229.4$ seconds. 
\begin{table}[htbp]
\vspace{-0.45cm}
\setlength{\abovecaptionskip}{-0cm}
\setlength{\belowcaptionskip}{-0cm}
\renewcommand{\arraystretch}{1.2}
\caption{Comparison of Computational Time between the MCS and the proposed PCE-based method}
\label{tab:WSCC9_time_comp}
\centering
\begin{tabular}{c|c|c|c|c}
\hline
Method & ${t_{\mathrm{ed}}(s)}$ & ${t_{\mathrm{pc}}(s)}$ & ${t_{\mathrm{es}}(s)}$ & ${t_{\mathrm{total}}(s)}$ \\
\hline
MCS    & --  &  --    & $2293591.51$
 & \bm{$2293591.51$}  \\
\hline
PCE   &  $7161.23$   & $4.76$  & $0.05$  &  \bm{$7166.04$} \\
\hline 
\end{tabular}
\vspace{-0.35cm}
\end{table} 

\vspace{-0.1cm}
\subsection{The PCCT Variation Control}
In \textbf{Step 5}, the first-order and the total Sobol' indices are calculated directly from the coefficients of the PCE-based model as shown in 
Table \ref{tab:Sobol_indices}, where $\chi_1$, $\chi_2$ and $\chi_3$ represent the random load power at bus 6, 8, and 5, respectively. Evidently, it can be seen 
that $\chi_2$ is the dominating random input that affects the the variance of PCCT. 
The small difference between $S_{\zeta}^{(1)}$ and $S_{\zeta}^{(T)}$ tells that the variance of PCCT is dominated by the variance of first-order terms (i.e. $g_{\bm{s}}(\chi_{\bm{s}}), \bm{s}=\lbrace\zeta\rbrace$) in \eqref{eq:SobolDecom}. It should be noted that such information may not be intuitive without detailed analysis. Indeed, both $\chi_2$ and $\chi_3$ are close to the fault as seen from Fig. \ref{fig:9BusSystem}, yet $\chi_2$ is the most sensitive random variable to the variation of PCCT according to the estimated Sobol' indices.  

To verify 
the impacts of $\chi_i$ on the variance of PCCT, 
the variance of  
each random input $\chi_i$ is smoothed to zero one by one, then the proposed PCE-based method is employed to predict probabilistic characteristics of PCCT 
(e.g., one of the three PCE-based models is built for $\chi_2$ with zero variance and $\chi_1, \chi_3$ with initial configurations in Section \ref{sec:Stastics_PCCT}). 
The results are presented in Table \ref{tab:smoothed_stat}. Clearly, smoothing $\chi_2$ is the most effective way to reduce the variance of PCCT ($\approx 70\%$ reduction) and increase the probability of stability ($\approx10\%$ increase) for the same FCT. Specifically, when the FCT is equal to 5 cycles, i.e., $0.083$s, the probability of stability is $88.99\%$ before smoothing any random variables, while it is $99.19\%$ after smoothing $\chi_2$.  
\begin{table}[htbp]
\vspace{-0.4cm}
\setlength{\abovecaptionskip}{-0cm}
\setlength{\belowcaptionskip}{-0cm}
\caption{First-Order and Total Sobol' Indices for all inputs}
\label{tab:Sobol_indices}
\centering
\begin{tabular}{c|c|c|c}
\hline
Sobol' Indices & ${\chi_1}$ & ${\chi_2}$ & ${\chi_3}$ \\
\hline
$S_{\zeta}^{(1)}$    & $0.0893$  &  $0.7080$    & $0.1947$  \\
\hline
$S_{\zeta}^{(T)}$  &  $0.0960$   & $0.7148$  & $0.1990$ \\
\hline 
\end{tabular}
\vspace{-0.3cm}
\end{table} 
%
%
\begin{table}[htbp]
\vspace{-0.4cm}
\setlength{\abovecaptionskip}{-0cm}
\setlength{\belowcaptionskip}{-0cm}
\renewcommand{\arraystretch}{1.2}
\caption{Statistics and Probability of Stability Predicted by PCE-based Method after Smoothing}
\label{tab:smoothed_stat}
\centering
\begin{tabular}{c|c|c|c|c}
\hline
 Index & ${\chi_1}$ & ${\chi_2}$ & ${\chi_3}$ & Before Smoothing \\
\hline
${\mu}$   & $0.1014$  &  $0.1019$    & $0.1013$ & $0.1011$ \\
\hline
${\sigma^2(10^{-4})}$  &  $1.9301$   & $\bm{0.6227}$  & $1.7460$ & $\bm{2.004}$ \\
\hline 
${P^{(S)}}$  &  $90.05\%$   & $\bm{99.19\%}$  & $92.23\%$ & $\bm{88.99\%}$ \\
\hline
\end{tabular}
\begin{tablenotes}
\item * $P^{(S)}$ denotes the probability of stability under preset FCT $=5$ cycles.
\end{tablenotes}
\vspace{-0.25cm}
\end{table} 

\begin{figure}[!t]
\centering
\includegraphics[width=0.75\columnwidth]{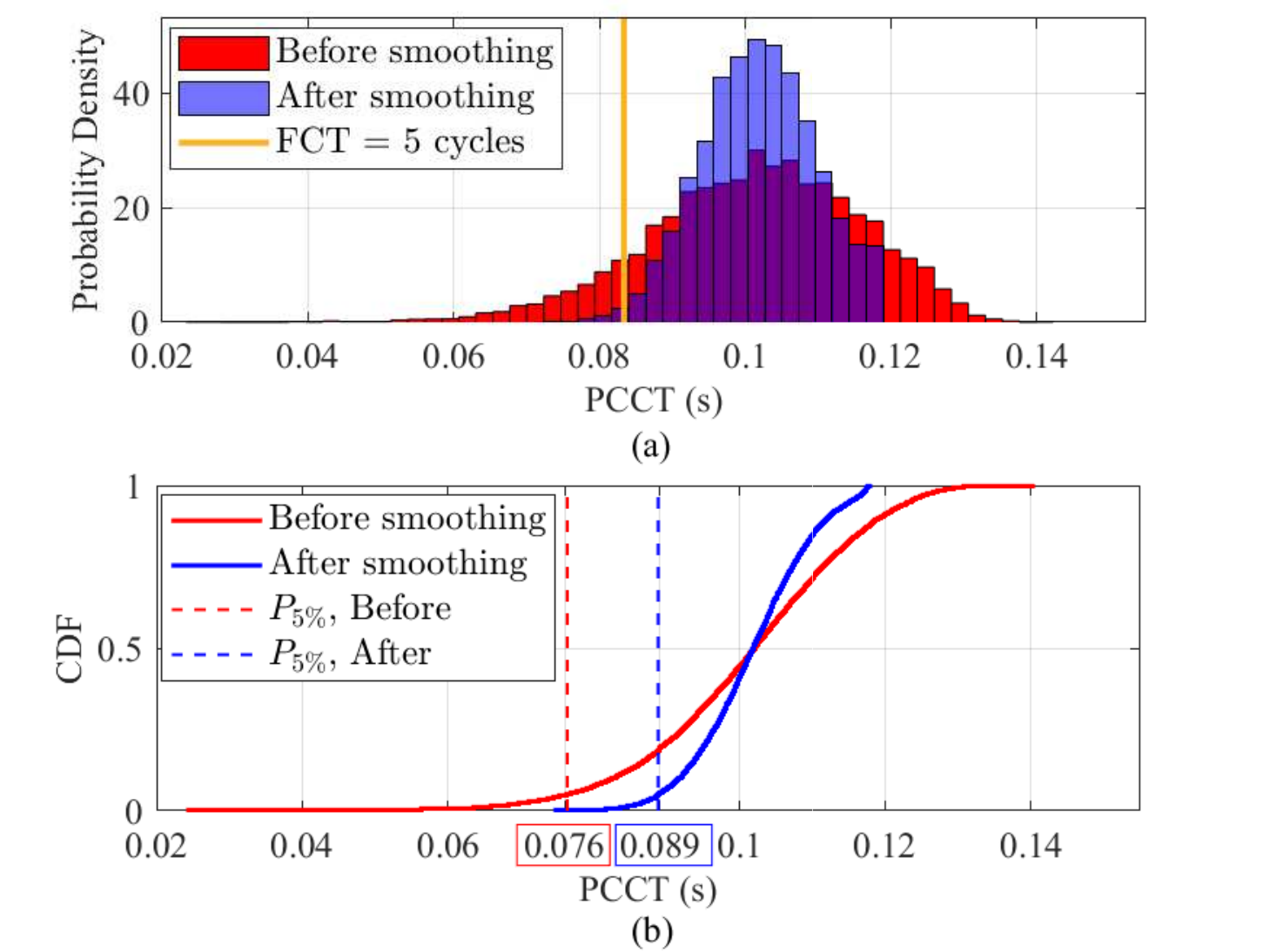}
\vspace{-0.2cm}
\caption{PDF and CDF comparison of PCCT before and after smoothing $\chi_2$. All the results are based on $10^4$ samples MCS. The “after smoothing” results are obtained by smoothing $\chi_2$ to its mean value. The red and blue dashed lines indicate the 5th percentiles $P_{5\%}$ before and after smoothing, respectively.}
\label{fig:PDFOfPCCT_BASmoothing_withQ}
\vspace{-0.6cm}
\end{figure}

Given the obtained sensitivity information, assume $\chi_2$ is smoothed out by, for example, an energy storage system. To verify the effectiveness of the mitigation measure, 
another set of $10^4$ samples TDS-based MCS with $\chi_2$ smoothed is executed.
The computation time comparison for this case is similar to Table \ref{tab:WSCC9_time_comp}.  Fig. \ref{fig:PDFOfPCCT_BASmoothing_withQ} provides  PDF and CDF comparisons before and after the smoothing,  
showing that smoothing $\chi_2$ reduces the variation of PCCT significantly. To maintain transient stability with $95\%$ confidence level, the FCT should be $0.076$s before the smoothing, while FCT can be extended to $0.089$s after smoothing out $\chi_2$. If FCT is preset to be $0.083$s, the probability that the system is stable after the fault is increased from $89.25\%$ to $99.23\%$. 

\section{Conclusions}
In this paper, an adaptive sparse PCE-based method has been proposed for 
probabilistic transient  stability assessment and enhancement. 
Simulation results show that the proposed method can provide accurate probabilistic and sensitivity information of PTTC efficiently. Moreover, smoothing the variance of the identified critical random input can significantly reduce the variance of PCCT and thus enhance the transient stability in the probability sense. 
Future work includes the application of the proposed method in large scale systems considering  
the dynamics and uncertainties of renewable energy sources.






\bibliographystyle{IEEEtran}
\bibliography{IEEEabrv,IEEEexample.bib}

\begin{thebibliography}{10}
\providecommand{\url}[1]{#1}
\csname url@samestyle\endcsname
\providecommand{\newblock}{\relax}
\providecommand{\bibinfo}[2]{#2}
\providecommand{\BIBentrySTDinterwordspacing}{\spaceskip=0pt\relax}
\providecommand{\BIBentryALTinterwordstretchfactor}{4}
\providecommand{\BIBentryALTinterwordspacing}{\spaceskip=\fontdimen2\font plus
\BIBentryALTinterwordstretchfactor\fontdimen3\font minus
  \fontdimen4\font\relax}
\providecommand{\BIBforeignlanguage}[2]{{%
\expandafter\ifx\csname l@#1\endcsname\relax
\typeout{** WARNING: IEEEtran.bst: No hyphenation pattern has been}%
\typeout{** loaded for the language `#1'. Using the pattern for}%
\typeout{** the default language instead.}%
\else
\language=\csname l@#1\endcsname
\fi
#2}}
\providecommand{\BIBdecl}{\relax}
\BIBdecl

\bibitem{faried2010}
S.~Faried, R.~Billinton, and S.~Aboreshaid, ``Probabilistic evaluation of
  transient stability of a power system incorporating wind farms,'' \emph{IET
  Renewable Power Generation}, vol.~4, no.~4, pp. 299--307, 2010.

\bibitem{2019Fortulan}
R.~L.~V. Fortulan and L.~F.~C. Alberto, ``Transient stability analysis of a
  single machine infinite bus system with uncertainties in generated power,''
  in \emph{2019 IEEE Power Energy Society General Meeting (PESGM)}, 2019, pp.
  1--5.

\bibitem{Roberts2015}
L.~G.~W. Roberts, A.~R. Champneys, K.~R.~W. Bell, and M.~di~Bernardo,
  ``Analytical approximations of critical clearing time for parametric analysis
  of power system transient stability,'' \emph{IEEE Journal on Emerging and
  Selected Topics in Circuits and Systems}, vol.~5, no.~3, pp. 465--476, 2015.

\bibitem{taj2018}
M.~Tajdinian, A.~R. Seifi, and M.~Allahbakhshi, ``Calculating probability
  density function of critical clearing time: novel formulation, implementation
  and application in probabilistic transient stability assessment,''
  \emph{International Journal of Electrical Power \& Energy Systems}, vol. 103,
  pp. 622--633, 2018.

\bibitem{Songkai2020}
S.~Liu, L.~Liu, Y.~Fan, L.~Zhang, Y.~Huang, T.~Zhang, J.~Cheng, L.~Wang,
  M.~Zhang, R.~Shi, and D.~Mao, ``An integrated scheme for online dynamic
  security assessment based on partial mutual information and iterated random
  forest,'' \emph{IEEE Transactions on Smart Grid}, vol.~11, no.~4, pp.
  3606--3619, 2020.

\bibitem{zhu2019}
L.~Zhu, D.~J. Hill, and C.~Lu, ``Hierarchical deep learning machine for power
  system online transient stability prediction,'' \emph{IEEE Transactions on
  Power Systems}, vol.~35, no.~3, pp. 2399--2411, 2019.

\bibitem{Sheng2018}
H.~Sheng and X.~Wang, ``Applying polynomial chaos expansion to assess
  probabilistic available delivery capability for distribution networks with
  renewables,'' \emph{IEEE Transactions on Power Systems}, vol.~33, no.~6, pp.
  6726--6735, 2018.

\bibitem{Wang2021}
X.~Wang, X.~Wang, H.~Sheng, and X.~Lin, ``A data-driven sparse polynomial chaos
  expansion method to assess probabilistic total transfer capability for power
  systems with renewables,'' \emph{IEEE Transactions on Power Systems},
  vol.~36, no.~3, pp. 2573--2583, 2021.

\bibitem{Wang2021ED}
X.~Wang, R.~Liu, X.~Wang, Y.~Hou, and F.~Bouffard, ``A data-driven uncertainty
  quantification method for stochastic economic dispatch,'' \emph{IEEE
  Transactions on Power Systems}, pp. 1--1, 2021.

\bibitem{Bingqing2019}
B.~Xia, H.~Wu, Y.~Qiu, B.~Lou, and Y.~Song, ``A galerkin method-based
  polynomial approximation for parametric problems in power system transient
  analysis,'' \emph{IEEE Transactions on Power Systems}, vol.~34, no.~2, pp.
  1620--1629, 2019.

\bibitem{Yijun2019}
Y.~Xu, L.~Mili, A.~Sandu, M.~R.~v. Spakovsky, and J.~Zhao, ``Propagating
  uncertainty in power system dynamic simulations using polynomial chaos,''
  \emph{IEEE Transactions on Power Systems}, vol.~34, no.~1, pp. 338--348,
  2019.

\bibitem{Miao2021}
M.~Fan, Z.~Li, T.~Ding, L.~Huang, F.~Dong, Z.~Ren, and C.~Liu, ``Uncertainty
  evaluation algorithm in power system dynamic analysis with correlated
  renewable energy sources,'' \emph{IEEE Transactions on Power Systems},
  vol.~36, no.~6, pp. 5602--5611, 2021.

\bibitem{wiener1938homogeneous}
N.~Wiener, ``The homogeneous chaos,'' \emph{American Journal of Mathematics},
  vol.~60, no.~4, pp. 897--936, 1938.

\bibitem{UQdoc_PCE}
S.~Marelli, N.~L\"uthen, and B.~Sudret, ``{UQLab user manual -- Polynomial
  chaos expansions},'' Chair of Risk, Safety and Uncertainty Quantification,
  ETH Zurich, Switzerland, Tech. Rep., 2021, report \# UQLab-V1.4-104.

\bibitem{Blatman2009}
G.~Blatman, ``Adaptive sparse polynomial chaos expansions for uncertainty
  propagation and sensitivity analysis,'' Ph.D. dissertation, Universite Blaise
  Pascal, Clermont-Ferrand, France, 2009.

\bibitem{Sobol1993}
S.~IM, ``Sensitivity estimates for nonlinear mathematical models,'' \emph{Math.
  Model. Comput. Exp}, vol.~1, no.~4, pp. 407--414, 1993.

\bibitem{UQdoc_Sen}
S.~Marelli, C.~Lamas, K.~Konakli, C.~Mylonas, P.~Wiederkehr, and B.~Sudret,
  ``{UQLab user manual -- Sensitivity analysis},'' Chair of Risk, Safety and
  Uncertainty Quantification, ETH Zurich, Switzerland, Tech. Rep., 2021, report
  \# UQLab-V1.4-106.

\bibitem{Milano2005}
F.~Milano, ``An open source power system analysis toolbox,'' \emph{IEEE
  Transactions on Power Systems}, vol.~20, no.~3, pp. 1199--1206, 2005.

\bibitem{marelli2014}
S.~Marelli and B.~Sudret, ``Uqlab: A framework for uncertainty quantification
  in matlab,'' in \emph{Vulnerability, uncertainty, and risk: quantification,
  mitigation, and management}, 2014, pp. 2554--2563.

\end{thebibliography}

\end{document}